# Accelerating PageRank Algorithmic Tasks with a new Programmable Hardware Architecture

Md Rownak Hossain Chowdhury, Mostafizur Rahman
Division of Energy, Matters and Systems, University of Missouri-Kansas City (UMKC)
Kansas City, MO, US
{rhctmc, rahmanmo} @umkc.edu

*Abstract*— Addressing the growing demands of artificial intelligence (AI) and data analytics requires new computing approaches. In this paper, we propose a reconfigurable hardware accelerator designed specifically for AI and data-intensive applications. Our architecture features a messaging-based intelligent computing scheme that allows for dynamic programming at runtime using a minimal instruction set. To assess our hardware's effectiveness, we conducted a case study in TSMC 28nm technology node. The simulation-based study involved analyzing a protein network using the computationally demanding PageRank algorithm. The results demonstrate that our hardware can analyze a 5,000-node protein network in just 213.6 milliseconds over 100 iterations. These outcomes signify the potential of our design to achieve cutting-edge performance in next-generation AI applications.

## I. INTRODUCTION

With the rapid growth of artificial intelligence (AI), researchers are increasingly leveraging its capabilities in fields like bioinformatics and genomics, aiming to decode complex biological processes at the molecular level [1]. Protein network analysis has become an essential method for mapping and understanding these processes, as it reveals the intricate molecular interactions that underpin cellular functions[2]. Moreover, as researchers curate larger, more comprehensive protein network datasets, such as hu.MAP 2.0[3], HuRI[4], DroRI etc.[5], AI-driven methodologies are becoming increasingly essential to uncover meaningful insights from this data. One particularly effective method in this context is the PageRank algorithm[6], initially designed for web ranking but now valued for its ability to capture both global and local network structures. This adaptability makes it especially suited for identifying key proteins and their roles within complex biological pathways[7], offering a powerful tool for advancing biomedical research.

Considering the complex nature of protein network and the immense growth in data volumes, computational demands have surged, placing significant strain on existing hardware architectures[8]. So far, general-purpose computers, particularly GPUs[9], have long been the workhorses for AI processing, excelling in parallel computation tasks. However, with the diminishing returns of Moore's law[10], there is a growing consensus on developing domain-specific specialized hardware solutions to continue improving computational efficiency[11]. Consequently, several domain-specific accelerators like SpArch[12], MatRaptor[13] have been proposed with improved dataflow mechanisms tailored for AI applications. Despite their strengths, these accelerator's fixed architectures and limited programmability restrict their adaptability to cater a wide range of AI implementations. As AI algorithms evolve rapidly, these rigid designs often lead to shorter lifecycles and increased nonrecurring engineering costs when adjustments to accommodate new models or tasks are required[14].

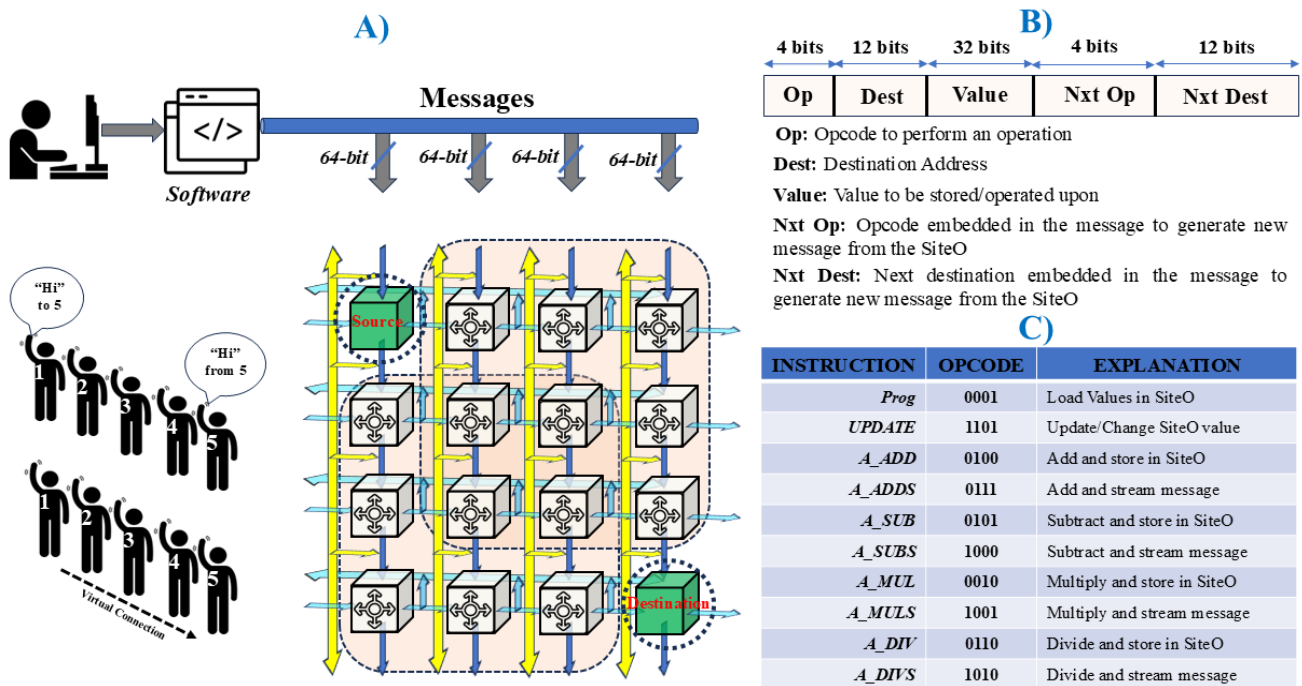

**Figure 1** (**A**) Overview and concept of programmable hardware architecture (**B**) Message encoding scheme (**C**) Instruction Set Architecture (ISA).

In response to these challenges, the computer architecture community aims to develop computing technologies that enable hardware to evolve alongside AI advancements without requiring constant redesigns[15]. Current and prior efforts towards reconfigurable AI accelerators primarily focus on FPGA [16] and Coarse Grain Reconfigurable Architectures (CGRAs) [17]. The deficiencies of FPGAs in terms of large overhead in interconnect and logic for mapping general purpose compute [18] have led to the developments of CGRAs. CGRA topologies vary widely in terms of core constructs, communication and system-on-chip integration[19][20]. Typically, CGRAs rely on an array of Processing Elements (PEs) interconnected via reconfigurable links, where the compiler primarily dictates execution flow, programming individual PEs and managing n-to-n routing between them. However, several challenges emerge, including a) limited runtime reconfigurability during task execution, b) complexity in managing registers and routing as systems scale, and c) inefficiencies due to separate instruction and data memory.

These limitations have motivated us to develop a messaging-based programmable hardware architecture designed specifically for AI and data-intensive tasks. Our approach diverges from traditional CGRAs by using intelligent processing elements that autonomously route data based on message content, without requiring explicit compiler intervention. Unlike CGRAs, our messaging-based architecture allows for dynamic message generation and runtime reconfigurability, leveraging the deterministic nature of AI tasks. By embedding both data and instructions in a single message, our architecture eliminates the need for separate instruction and data memories, significantly reducing the network-on-chip (NoC) overhead and complexity associated with n-to-n routing. This allows our hardware to adjust data flow dynamically, enhancing its adaptability and real-time responsiveness compared to CGRAs' fixed task configurations during execution. Therefore, the hardware architecture can handle large amount of data with the same efficiency since the reconfiguration and computation scheme is not rigid for certain data sizes.

In short, the key contributions of our design are as follows:
- An innovative computer architecture concept to achieve programmability at run-time.
- A compact instruction set architecture that enables intuitive data flow mechanism within hardware elements without software intervention.
- A low-latency matrix-vector multiplication approach using reconfigurable hardware structure.
- Evaluation of our programmable accelerator on computation-intensive PageRank algorithm to analyze protein network with high throughput.

II. HARDWARE ARCHITECTURE

A. *Programmability*

Our innovation is a programmable computing architecture that can be reconfigured at run-time to behave as a custom ASIC through interconnection flexibility and, as a result, information processing. An analogy of the proposed interconnect configurability is shown in Figure 1A (left). If we assume that 5 people are standing in a line, and they generate and pass messages from left to right in a circular manner (#1 sends to #2, #2 sends to #3, and the rightmost person, #5, sends its message to #1), then any message from anyone can be delivered to anyone in this human chain. For example, if #1 generates a message "Hi" intended for #5, it passes it to #2, and #2, #3, and #4 keep passing the same message to their right until destination #5 is reached. In this message passing scheme, #2, #3, and #4 form a virtual link between #1 and #5. This source, destination-based message passing scheme can be applied to computing cores as well as shown in Figure 1A(right). We organize the computational units in rows and columns and connect them in a manner that messages can be communicated between any units. The hardware units are very light and capable of performing only essential operations like programming and arithmetic. A message originating from any of the units in a 16 unit configuration (4 rows and 4 columns- Figure 1A(right)), goes right or down depending on the destination. To configure our proposed hardware unit, only messages need to be sent to proper units which in turn sets destination addresses, values, and operations through register writes. The message encoding scheme is shown in Figure 1B. A computational site is capable of both programming and arithmetic operations. The opcode values act as guides to distinguish between programming and operation. The instruction set architecture (*ISA*) of our programmable hardware accelerator is shown in Figure 1C, that comprises only *10* instructions. Among these, one instruction (*Prog*) is designated for data loading and the remaining nine instructions (*UPDATE, A_ADD, A_SUB, A_MUL, A_DIV, A_ADDS, A_SUBS, A_MULS, A_DIVS*) are allocated for various mathematical manipulation.

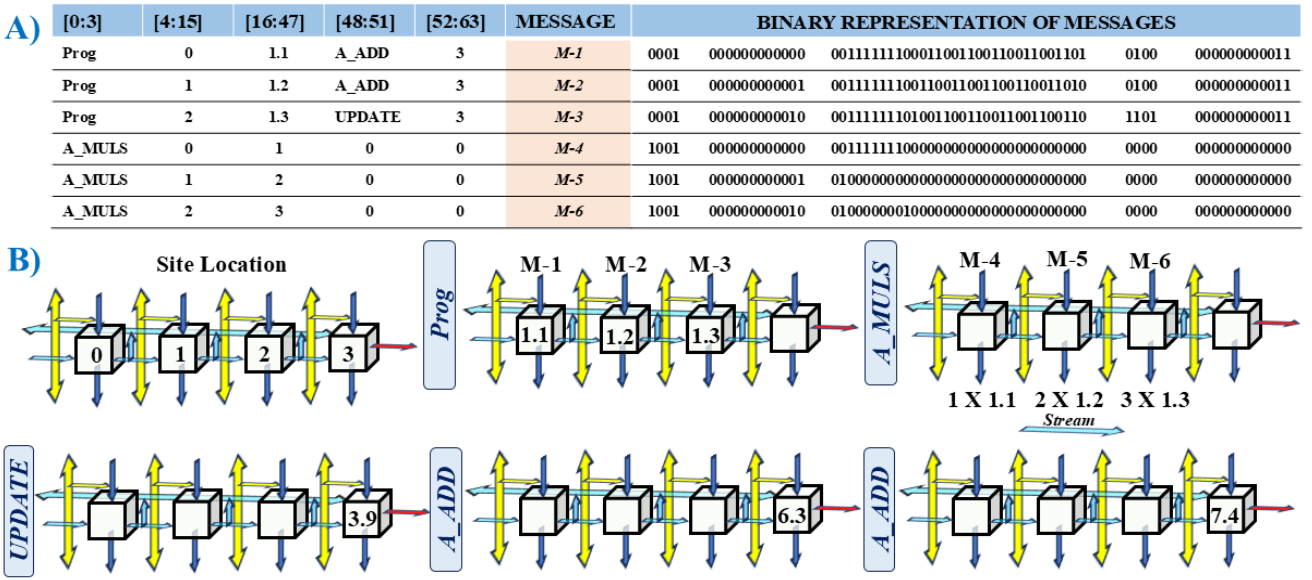

**Figure 2** (**A**) Bitwise segmentation of messages and (**B**) Illustration of the Concept of programmability utilizing hardware resources.

The programmability concept of our proposed hardware accelerator is illustrated in Figure 2. Here, we used only four instructions and focused solely on a single row of the entire architecture for ease of demonstration. In this example, six messages shown in Figure 2A are provided as inputs to three sites (Site0, Site1, Site2) over two separate clock cycles. During the initial clock cycle, three messages (*M-1*, *M-2*, and *M-3*) are transmitted from the user end. As soon as, sites acknowledge these messages, the decoder unit of each site analyze the opcode (*from bit 0 to 3*) and the destination location (*from bit 4 to 15*). In this case, the opcode is "*Prog*" and hence the values (*1.1*, *1.2*, *1.3*) embedded in the message will be stored in the floating-point unit (FPU) of respective sites. Sites also retain the next opcode (*from bit 48 to 51*) and the next destination (*from bit 52 to 63*) integrated in the message. In the next clock cycle, the remaining three messages (*M-4*, *M-5*, and *M-6*) are sent. According to the opcode (*A_MULS*) of these messages, it first multiplies values (*1*, *2*, *3*) contained in the messages with the values (*1.1*, *1.2*, *1.3*) stored in the respective FPU. The opcode and destination are then updated according to the next opcode and next destination value stored in the site. Consequently, three multiplication results (*1.1*, *2.4*, *3.9*) will be streamed towards site3 with opcode "*A_ADD*", "*A_ADD*", and "*UPDATE*" respectively. Thus, site3 updates its value to *3.9* at first and then performs two consecutive addition operations. Finally, it stores *7.9* in site3.

### B. Matrix-Vector Multiplication

The matrix-vector multiplication operation using our proposed hardware unit can be divided into four (4) steps: *1*) Data (*Matrix*) load *2*) Data (*Vector*) load and multiply and *3*) Addition and *4*) Offload. Figure 3 illustrates an example of the matrix-vector multiplication between *matrix A* (*4 X 3*) and *vector B* (*3 X 1*), showing the required time steps in each stage. Firstly, the entire *matrix A* is distributed across the fabric through hopping. The decoder unit of each site decodes the incoming message and programs each site with proper values. Upon loading *matrix A*, the transpose of the *vector B* is transferred leveraging vertical bus, and multiplication operations are performed in each site. Afterward, the multiplication results are streamed to the desired location using horizontal bus, upgrading the message. Finally, the matrix-vector multiplication results are stored in the last site of each row of the hardware architecture after executing addition.

Suppose a matrix-vector multiplication takes two inputs: a matrix of size (*N X M*) and a vector of size (*M X 1*) then the number of sites required to store a matrix is (*N X M*). Since the vector has only *1* column, it will not occupy any additional sites. Additionally, we need another *N* number of sites to add the multiplication result. Thus, the number of sites necessary to perform a matrix-vector multiplication is *{(N X M) + N}*. Here, the matrix will be loaded through hopping where each row of the matrix will be transferred one by one starting from the last row. Hence, the number of time steps required to load a matrix of size (*N X M*) is *N*. Then, the vector will be sent using vertical bus and it will be multiplied with the matrix. This operation will require only *1* time step because of using vertical bus. Similarly, the addition operation will cost *1* time step utilizing horizontal bus. Finally, the result will be offloaded that will consume *1* more step. Thus, the total time steps required for the matrix-vector multiplication operation is (*N + 3*).

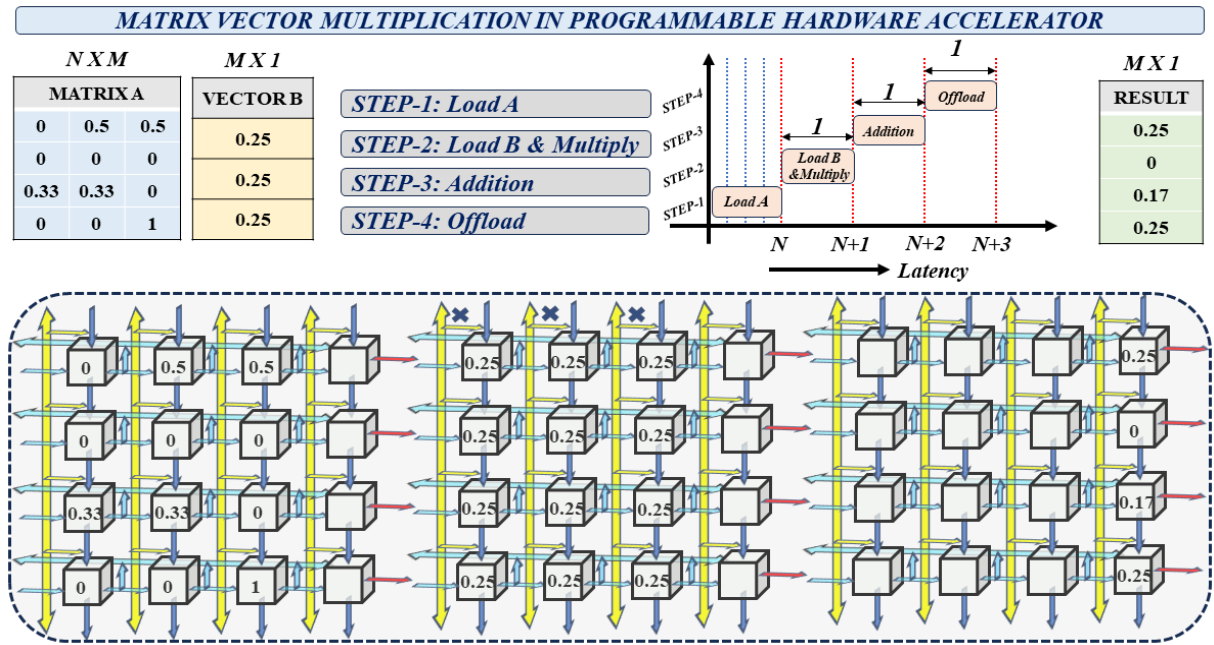

**Figure 3** Matrix-Vector multiplication in our proposed programmable hardware accelerator.

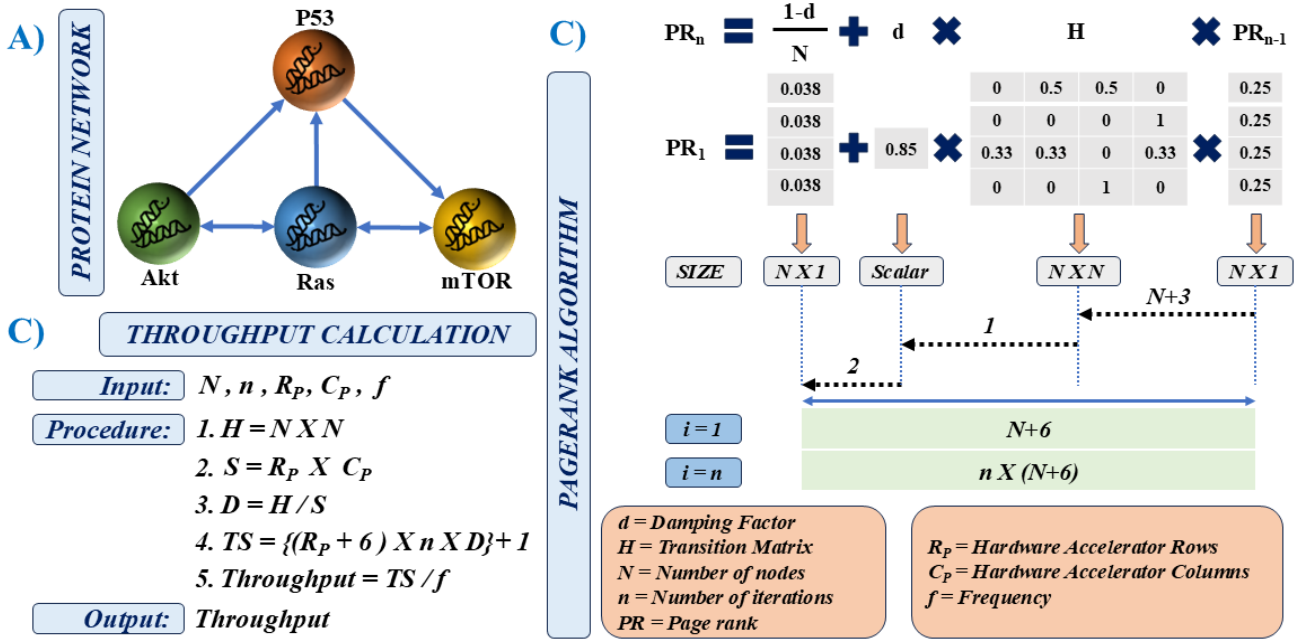

**Figure 4** (**A**) A small portion of large Protein network (**B**) Time steps required to implement the PageRank algorithm using our proposed hardware accelerator (**C**) Throughput calculation of the PageRank algorithm utilizing available resources.

### III. ACCELERATING PROTEIN NETWORK ANLYSIS TASK

While analyzing protein network using PageRank algorithm, each protein is defined as nodes and the connections between proteins are referred to as edges. The PageRank algorithm determines the importance of proteins (nodes) by analyzing their connections (edges). Each protein's significance is iteratively updated based on its interactions with neighboring proteins. The algorithm follows the equation shown in Figure 4(B), which includes key parameters like the damping factor (d), the transition matrix (H), and the rank vector from the previous iteration ($PR_{n-1}$). This iterative process ultimately calculates the rank vector ($PR_n$), representing the relative importance of each protein in the network. Therefore, interpreting data-intensive protein network using PageRank algorithm puts forward various computational challenges to handle vast amount of data as manipulating such large datasets involves multiple iterations and computation expensive operation like matrix-vector multiplication. In such scenario, our programmable hardware architecture offers a convenient solution by accelerating the matrix-vector multiplication process.

To showcase the performance of our programmable hardware design, we have segmented the PageRank algorithm in several stages and calculated the required time steps in every stage over many iterations. Figure 4(B) demonstrates the total time steps required to complete one iteration of the PageRank algorithm. Here, the multiplication between the matrix (*H*) and the vector ($PR_{n-1}$) requires *N+3* steps; then, a scalar (*d*) will be loaded to be multiplied with the result of matrix vector multiplication that costs *1* time step. After that, the addition and offloading require *2* more steps. This process will be repeated several times based on the desired accuracy. Thus, the total time steps necessary for *n* iterations can be presented as: {*n X (N+6)*}; where, *N* represents number of proteins in the network and *n* represents number of iterations. Based on this concept, the throughput of a PageRank algorithm for a large dataset using limited hardware resources can be calculated as shown in Figure 4(C).

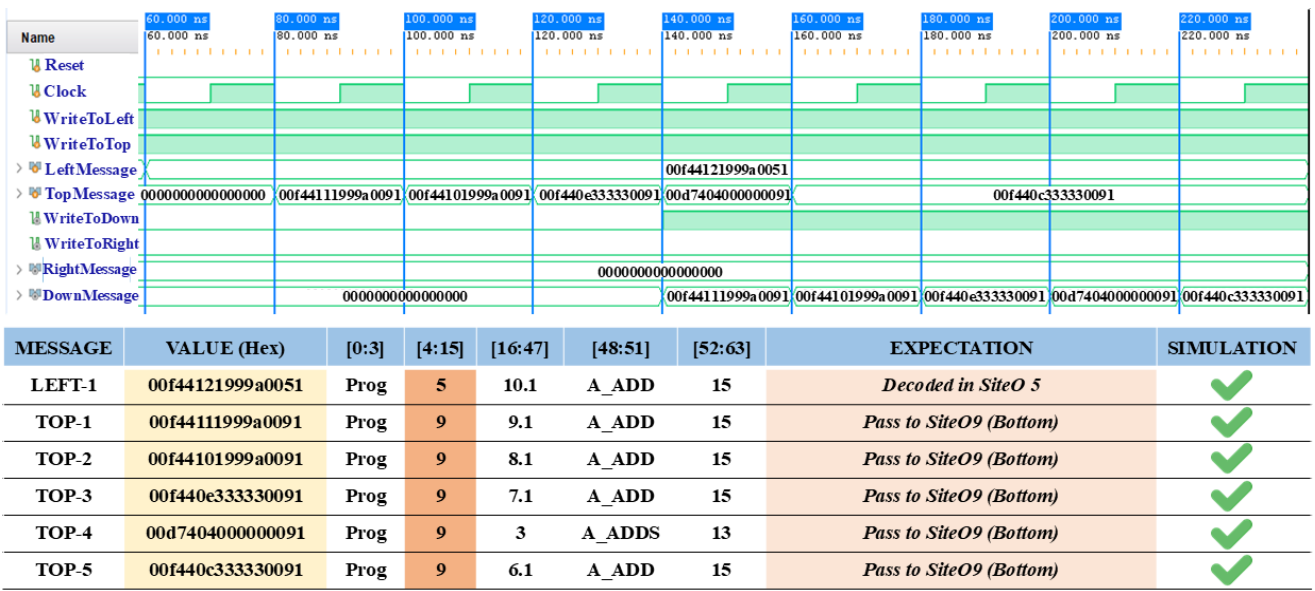

| MESSAGE | VALUE (Hex) | [0:3] | [4:15] | [16:47] | [48:51] | [52:63] | EXPECTATION | SIMULATION |
|---|---|---|---|---|---|---|---|---|
| LEFT-1 | 00f44121999a0051 | Prog | 5 | 10.1 | A_ADD | 15 | Decoded in SiteO 5 | ✓ |
| TOP-1 | 00f44111999a0091 | Prog | 9 | 9.1 | A_ADD | 15 | Pass to SiteO9 (Bottom) | ✓ |
| TOP-2 | 00f44101999a0091 | Prog | 9 | 8.1 | A_ADD | 15 | Pass to SiteO9 (Bottom) | ✓ |
| TOP-3 | 00f440e333330091 | Prog | 9 | 7.1 | A_ADD | 15 | Pass to SiteO9 (Bottom) | ✓ |
| TOP-4 | 00d7404000000091 | Prog | 9 | 3 | A_ADDS | 13 | Pass to SiteO9 (Bottom) | ✓ |
| TOP-5 | 00f440c333330091 | Prog | 9 | 6.1 | A_ADD | 15 | Pass to SiteO9 (Bottom) | ✓ |

**Figure 5** Simulation result of our proposed reconfigurable hardware architecture.

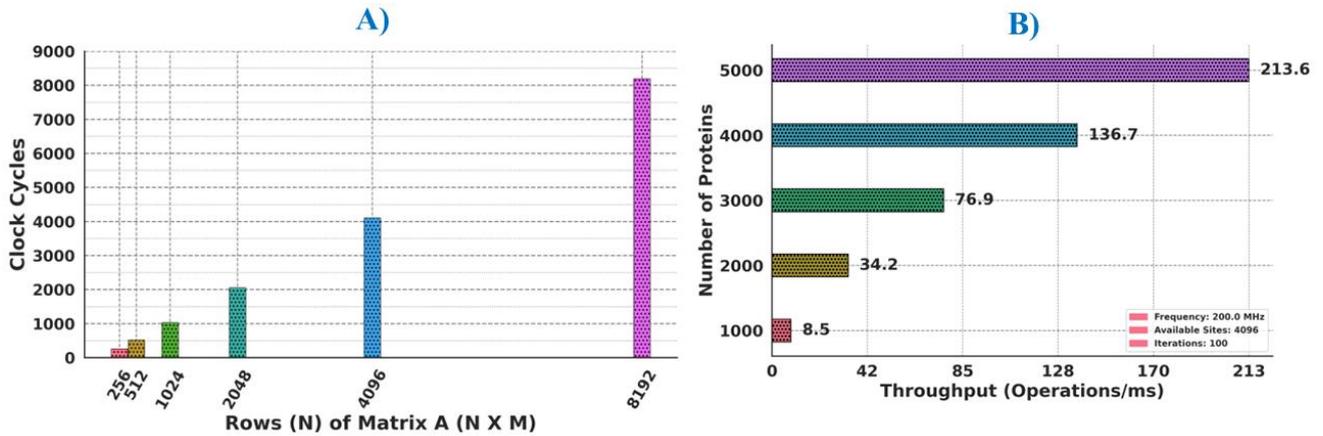

**Figure 6** (**A**) Latency calculation of Matrix-Vector multiplication varying rows of Matrix A (**B**) Throughput calculation varying the number of proteins in protein network.

*A. Methodology*

we developed our programmable hardware architecture using Verilog and simulated the design in Xilinx Vivado to verify the desired functionality of our design. Later, we followed a digital design flow based on CAD tools using a high-performance compact mobile computing plus (*CLN28HPC+*) process from TSMC 28nm commercial PDK with 8 metal layers and supply voltage of 0.9V. We primarily focused on demonstrating programmable nature of our hardware to validate our novel computing concept and measured several design metrics to examine the hardware implementation benefits. Later, we used our computing strategy to handle complex operation of PageRank algorithm to analyze protein networks and observed throughput with increasing network complexity. We have maintained a uniform *200* MHz clock frequency in various stages of implementation like simulation, synthesis, and performance analysis. The simulation result of our proposed hardware architecture is depicted in Figure5. In this testbench, we have considered six (*6*) messages (Left-1, TOP-1 to TOP-5) to be sent to our hardware unit. The current address of our hardware unit is *5*, while the addresses of neighboring hardware units are *2* (Top), *9* (Bottom), Left (*4*), Right (*6*). Here, the opcode and the current destination of the left message-1 are "*Prog*" and "*5*" respectively; hence, the message will be decoded inside the SiteO rather than passing it to the right or bottom. On the other hand, the opcode and the present location of messages that are sent from the top side (TOP-1 to TOP-5) are "*Prog*" and "*9*". Therefore, these messages should be routed through the bottom port. The expected behavior in terms of programmability is achieved which is evident in simulation result shown in Figure 4. A single unit of our programmable architecture consumes only *4.1* mW power under TSMC *28nm* technology as reported in Table I.

*B. Performance*

The latency of our matrix-vector multiplication process is appended in Figure 6 (A). Here, we have varied the number of rows (*N*) of the matrix of size (*N X M*) from 256 to 8192 and observed the respective latency. The results indicate that the time steps required to complete a matrix-vector multiplication is almost equal (*N+3* ≈ *N*) to the number of rows in the matrix and it is independent of the number of columns in the matrix or the vector size. Figure 6 (B) demonstrates the throughput to analyze protein network utilizing PageRank algorithm. In this case, we have varied the number of proteins in the network from 1000 to 5000 and observed required times. For evaluation, we have conducted 100 iterations utilizing a *200* MHz clock frequency and leveraging only *4096* available hardware units. Our proposed computing methodology requires just *213.6* milliseconds to complete *100* iterations of PageRank algorithm to analyze a protein network comprising 5000 proteins.

**Table I:**
**Design parameters of a single programmable site**

| Technology | TSMC 28nm |
|---|---|
| Process | HPC+ |
| Metal Layer | 1P8M |
| Voltage (VDD) | 0.9V (nominal) |
| Package | Wire Bond |
| Area | 6 mm$^2$ |
| Power | 4.1 mW |
| Frequency | 200 MHz |
| Gate Count | ~98000 |

IV. CONCLUSION

We presented a novel configurable hardware architecture designed for AI and data-intensive tasks. This architecture leverages a flexible interconnection scheme and a parallel organization of compute units, making it well-suited for large-scale, parallel, computationally intensive tasks. Additionally, we introduced the associated instruction set architecture (ISA) for programming and operation control. To demonstrate the efficacy of our hardware, we presented a case study involving a protein network search. Furthermore, we validated our design concepts through simulations conducted on a TSMC 28nm technology node. We also detailed our evaluation methodology and presented performance results for running the matrix-vector multiplication tasks of the PageRank algorithm within the context of protein network search. The combination of our innovative runtime-programmable architecture for compute-intensive tasks and its demonstrably fast performance suggests its significant potential for future applications.

## V. REFERENCES

[1] N. Sapoval et al., "Current progress and open challenges for applying deep learning across the biosciences," *Nature Communications*, vol. 13, no. 1, Apr. 2022, **DOI: 10.1038/s41467-022-29268-7**

[2] I. A. Kovács et al., "Network-based prediction of protein interactions," *Nature Communications*, vol. 10, no. 1, p. 1240, Mar. 2019, **DOI: 10.1038/s41467-019-09177-y**

[3] K. Drew, J. B. Wallingford, and E. M. Marcotte, "hu.MAP 2.0: integration of over 15,000 proteomic experiments builds a global compendium of human multiprotein assemblies," *Molecular Systems Biology*, vol. 17, no. 5, May 2021, **DOI: 10.15252/msb.202010016**

[4] K. Luck et al., "A reference map of the human binary protein interactome," *Nature*, vol. 580, no. 7803, pp. 402–408, Apr. 2020, **DOI: 10.1038/s41586-020-2188-x**

[5] H.-W. Tang et al., "Next-generation large-scale binary protein interaction network for Drosophila melanogaster," *Nature Communications*, vol. 14, no. 1, p. 2162, Apr. 2023, **DOI: 10.1038/s41467-023-37876-0**

[6] S. Brin and L. Page, "The anatomy of a large-scale hypertextual Web search engine," *Computer Networks and ISDN Systems*, vol. 30, no. 1–7, pp. 107–117, Apr. 1998, **DOI: 10.1016/s0169-7552(98)00110-x**

[7] X. Lei, J. Liang, and L. Guo, "Identify protein complexes based on PageRank algorithm and architecture on dynamic PPI networks," *International Journal of Data Mining and Bioinformatics*, vol. 22, no. 1, p. 350, 2019, **DOI: 10.1504/ijdmb.2019.101394**

[8] R. Hameed et al., "Understanding sources of inefficiency in general-purpose chips," *ACM SIGARCH Computer Architecture News*, vol. 38, no. 3, pp. 37–47, Jun. 2010, **DOI: 10.1145/1816038.1815968**

[9] W. J. Dally, S. W. Keckler and D. B. Kirk, "Evolution of the Graphics Processing Unit (GPU)," in *IEEE Micro*, vol. 41, no. 6, pp. 42-51, 1 Nov.-Dec. 2021, **DOI: 10.1109/MM.2021.3113475**

[10] T. N. Theis and H. .-S. P. Wong, "The End of Moore's Law: A New Beginning for Information Technology," in *Computing in Science & Engineering*, vol. 19, no. 2, pp. 41-50, Mar.-Apr. 2017, **DOI: 10.1109/MCSE.2017.29**

[11] J. L. Hennessy and D. A. Patterson, "A new golden age for computer architecture: Domain-specific hardware/software co-design, enhanced security, open instruction sets, and agile chip development," *2018 ACM/IEEE 45th Annual International Symposium on Computer Architecture (ISCA)*, Los Angeles, CA, USA, 2018, pp. 27-29, **DOI: 10.1109/ISCA.2018.00011**

[12] Z. Zhang, H. Wang, S. Han and W. J. Dally, "SpArch: Efficient Architecture for Sparse Matrix Multiplication," 2020 IEEE International Symposium on High Performance Computer Architecture (HPCA), San Diego, CA, USA, 2020, pp. 261-274, **DOI: 10.1109/HPCA47549.2020.00030**

[13] 33. N. Srivastava, H. Jin, J. Liu, D. Albonesi and Z. Zhang, "MatRaptor: A Sparse-Sparse Matrix Multiplication Accelerator Based on Row-Wise Product," *2020 53rd Annual IEEE/ACM International Symposium on Microarchitecture (MICRO)*, Athens, Greece, 2020, pp. 766-780, **DOI: 10.1109/MICRO50266.2020.00068**

[14] T. Nowatzki, V. Gangadhan, K. Sankaralingam and G. Wright, "Pushing the limits of accelerator efficiency while retaining programmability," *2016 IEEE International Symposium on High Performance Computer Architecture (HPCA)*, Barcelona, Spain, 2016, pp. 27-39, **DOI: 10.1109/HPCA.2016.7446051**

[15] T. Fujii et al., "New Generation Dynamically Reconfigurable Processor Technology for Accelerating Embedded AI Applications," *2018 IEEE Symposium on VLSI Circuits*, Honolulu, HI, USA, 2018, pp. 41-42, **DOI: 10.1109/VLSIC.2018.8502438**

[16] I. Kuon, R. Tessier, and J. Rose, "FPGA Architecture: Survey and Challenges," *Foundations and Trends® in Electronic Design Automation*, vol. 2, no. 2, pp. 135–253, 2007, **DOI: 10.1561/1000000005**

[17] Z. Li, D. Wijerathne, and T. Mitra, "Coarse Grained Reconfigurable Array (CGRA)." Available: **https://www.comp.nus.edu.sg/~tulika/CGRA-Survey.pdf**

[18] A. Podobas, K. Sano and S. Matsuoka, "A Survey on Coarse-Grained Reconfigurable Architectures from a Performance Perspective," in *IEEE Access*, vol. 8, pp. 146719-146743, 2020, **DOI: 10.1109/ACCESS.2020.3012084**

[19] L. Liu et al., "A Survey of Coarse-Grained Reconfigurable Architecture and Design," *ACM Computing Surveys*, vol. 52, no. 6, pp. 1–39, Oct. 2019, **DOI: 10.1145/3357375**

[20] M. Wijtvliet, L. Waeijen and H. Corporaal, "Coarse grained reconfigurable architectures in the past 25 years: Overview and classification," *2016 International Conference on Embedded Computer Systems: Architectures, Modeling and Simulation (SAMOS)*, Agios Konstantinos, Greece, 2016, pp. 235-244, **DOI: 10.1109/SAMOS.2016.7818353**